# Proposed benchmarks for PRT networks simulation

Jerzy Mieścicki[*]
Wiktor B. Daszczuk[**]

**Abstract**

Personal Rapid Transit (PRT) is a promising form of urban transport. Its operation consists in the use of small unmanned vehicles which convey the passengers among stations within a dedicated network. Various aspects of the PRT network performance are frequently evaluated using the discrete-event simulation. The paper supports the need of establishing some reference models for the simulation of PRT networks, targeted mainly towards the needs of the research on network management algorithms. Three models of such PRT network models are proposed and discussed. The presented models can play the role of benchmarks which would be very useful for comparative evaluation of heuristic control algorithms, developed by different research groups.

# 1. Introduction

Personal Rapid Transit (PRT) is a relatively new and promising form of urban transport. Its operation is based on the use of small electric vehicles (with the capacity of 4 or 6 persons each) conveying the passengers among stations within a dedicated network. The PRT system is *personal*, as the vehicles are used to satisfy the personal needs of passengers in a sense that vehicles are called on demand and they ride to specific destination, defined by the user at the beginning of the travel. It is also *rapid*, because the system of tracks is separated from the conventional traffic (e.g. elevated above the street level), which allows to increase velocity and enhance the safety of operation. Moreover, as the vehicles are identical, the traffic is pretty smooth and the emergence of traffic congestions and jams is greatly reduced.

The PRT vehicles are *unmanned*, that is, they do not need any human control during their regular operation. All the activities of the vehicle (including driving down the tracks, preserving separation between vehicles, stopping at stations, determining the optimal route from the given origin to the destination, responding to new passengers' calls etc.) are controlled by the appropriate real-time control algorithms. For the communication among vehicles as well as with the elements of the network infrastructure (stations, intersections, capacitors etc.) the wireless communication system is used. Therefore, the control structure of PRT is a complex, distributed, real-time system.

---

[*] Institute of Computer Science, Warsaw University of Technology, Warsaw, Poland, jms@ii.pw.edu.pl
[**] Institute of Computer Science, Warsaw University of Technology, Warsaw, Poland, wbd@ii.pw.edu.pl

The infrastructure of the network consists of stations and capacitors connected with unidirectional tracks or guideways. The track can split into two lanes (making the so-called *fork* intersection), similarly two lanes can merge into one track (*join* intersection). Functionally, one can consider the PRT network as a directed graph, where stations, capacitors and intersections (of both types) are *nodes* while segments of tracks are *edges*. This way the complex network of a rich topography can be deployed, covering the part of the city, airport, shopping centers, exhibition area etc.

The idea of PRT is not new. It can be dated back to 1970's [12]. However, neither a research done on that time [9, 12] nor a few practical implementations [2, 14, 15] resulted in a significant proliferation of PRT. Nowadays, however, the wave of interest in PRT returns and apparently grows, mainly due to the progress in telecommunications and computer technology (hardware as well as software) made in the last few decades. In contrast to 1970's, the contemporary technological solutions (available on the *off-the-shelf* basis) make the implementation of PRT systems more feasible. Several new PRT systems have been actually built [2, 14, 15], while dozens of other ones are planned or at least studied [1, 7, 9, 10], commissioned by e.g. local city authorities, management of airports etc. Still others are subjects of more theoretical research projects [e.g. 3, 4, 5,6].

## 2. The role of benchmarks (reference models) in the simulation of PRT networks

Several types of simulation models and tools are intensively used during the design of elements of the PRT network. Some of them are used in order to solve the detailed engineering problems of physical construction of tracks, dynamic properties of vehicles, etc. These issues are not specific to PRT and refer to general simulation methods in mechanical, electrical or construction technology. However, there are also numerous problems solved using the *network level simulation*, where the PRT network is viewed as a whole and its specific properties are broadly taken into account. In this case, the Discrete Event Simulation (DES, [7, 13]) is typically used.

In terms of Systems Engineering [11] we may say that the network level simulation is used as a tool for setting a balance between the needs between several groups of *stakeholders* involved in the implementation of PRT system. Typically, among the stakeholders of the PRT project are: Investor, Designer, Builder, Operator, Maintenance Provider and Customer. The viewpoints of these parts of the game are frequently conflicting. For instance, the large scope of the network or the large number of stations are desirable from the Customer's point of view while they are not necessarily favorable from the viewpoint of the Investor and Maintenance Provider. Similarly, small number of vehicles could reduce the network deployment cost but it can result in an increase of waiting time so that it is inacceptable to the Customer etc. The simulation on the network level helps to find out the balanced solutions acceptable to several conflicting counterparts.

The analysis of various design decisions and their impact on the network's cost and performance is usually the subject of a detailed report commissioned by some local authority [1, 8, 10]. Understandably enough, also the simulation model often addresses the specific, local conditions and requirements: the particular network's topography, the placement of stations and capacitors, specific demand characteristics etc. Therefore, the results of such an analysis can hardly be compared with other ones, maybe similar to an extent but based on different goals, requirements and terrain characteristics.

On the other hand, PRT networks are also subject of more theoretical research, unrelated to specific requirements of a given city or area. Among the most important research topics are the functional properties and effectiveness of network control algorithms. They play the crucial role as the vehicles are unmanned and their movement within the network is personalized according to the passengers' request.

The control algorithms can be subdivided into two classes: *coordination* and *management* algorithms. The former are designed to provide the secure movement of individual vehicles down the track, e.g. to preserve the assumed separation between vehicles, to safely solve the conflicts in intersections and to provide the proper behavior of a vehicle approaching and leaving the station berth. In order to investigate and test this class of algorithms a relatively simple simulation model would do. One or two segments of tracks, a few vehicles, one or two intersections, a station with a few berths etc. are only needed to test the algorithm for validity and effectiveness. The results rather easily scale up to larger networks, as the valid coordination algorithms designed for one intersection or one station are not dependent on network topography and preserve their validity in a network containing many intersections or stations.

The network *management* algorithms deal with other aspects of the network operation, mainly: (a) dynamic routing and (b) the empty vehicle management. The former problem consists in finding the optimal route between the origin and destination taking into consideration the changing traffic conditions. The latter deals with vehicles which are not actively used by passengers at a given moment. Empty vehicles can passively wait in stations where they were leaved after completing the last trip, but they have to be sent elsewhere if the berth must be freed. Also, they can be called from other stations whenever there is no empty vehicle available for new arriving passengers etc. Therefore, the whole fleet of vehicles is continuously managed by a set of sophisticated algorithms, involving also the possible prediction where the vehicles would be needed in the future.

The control algorithms are in general heuristic. Their performance (regardless of what performance indices are defined) can be evaluated mainly by collecting and comparing the numerical results from simulation. Unfortunately, in contrast to lower-level or coordination problems, the performance of management algorithms strongly depends on the topography and size of a given network and its multiple parameters (e.g. the number of stations and vehicles, number of berths in a station, distances among stations or intersections etc.). This greatly hinders comparative evaluation of different solutions known from the literature. Authors usually rely on the simulation of network models tailored to the particular topography of the site and objectives set by the authority contracting the design. They use different network topographies, vehicle and tracks parameters, different behavioral rules etc., so that the comparative analysis of results is hardly possible.

To allow a relatively objective comparison of different heuristic algorithms we should specify a set of standard or reference models of PRT networks along with the basic model parameters. Only then can a comparative analysis satisfy the important methodological principle of comparing *caeteris paribus*. The reference models (provided that they are widely known and accepted) can play the role of *benchmarks* in the research on control algorithms for PRT networks. The use of benchmarks is a common practice and has proved to be very fruitful in many disciplines of engineering, management science etc.

Below we present three simplified PRT network models that can be candidates for such reference models or benchmarks. We have used them in our research on simulation of PRT networks [3, 4, 5, 6].

# 3. Proposed PRT reference models

## 3.1. General rules

We have defined three models of different topography and a set of parameters and properties common to all of them. The models are referred to as *City*, *SeaShore* and *TwinCity*. Their mnemonic names correspond to the hypothetical environments they illustrate. In every case the network layout is, of course, simplified and symmetric. The track systems contain both short and long segments as well as both double track highways and unidirectional ordinary lines of lower speed limit. Number of stations is of an order of 10 or 12 etc. In other words, the networks are of moderate size and every model described below can be rather easily implemented in any simulation tool, just in order to analyze how the various network management strategies, multiple parameters of control algorithms etc. affect the network performance and effectiveness.

## 3.2. Three models of network topography

### 3.2.1. City

The first model, called *City* (Fig. 1), schematically reflects the structure of a traditional town. It consists of the downtown and eight peripheral housing areas. Four stations (A to D) serve the central zone (of 3 km diameter) while the remaining ones (E to L) are located in suburbs.

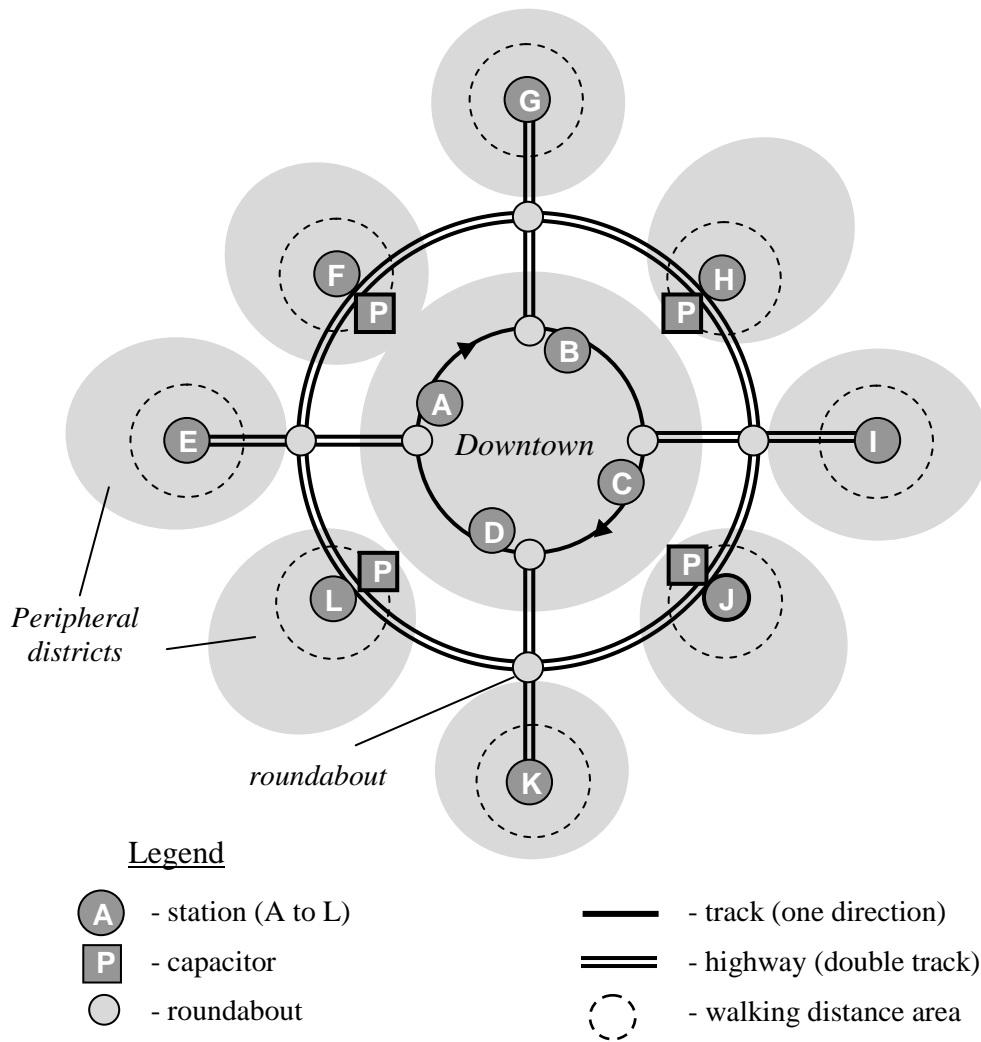

*Fig.* 1. *City* model

Internal structure of every station logically corresponds to the one schematically shown in Fig. 2. The passengers board the vehicles and alight them using several berths. Station has the by-pass connection or a segment of track allowing the vehicles travelling to other stations to omit the berths. Additionally, there are two buffers inside a station: entry buffer for incoming vehicles waiting for a free berth and exit buffer for vehicles leaving the station and temporarily waiting before they join the traffic.

It should be emphasized that the detailed internal structure of stations does *not* have to be modeled explicitly, as in Fig. 2, i.e. in terms of a graph consisting of berths, track segments fork/join intersections etc. The detailed behavior inside the station (selection of the free berth, docking at the berths, joining the traffic etc.) is controlled by the coordination level algorithms while our reference models are oriented towards the network-level simulation. Therefore, for the purpose of evaluation of management algorithms we can assume that every station is just an elementary node of the network. Its internal structure and the logic of operation is encapsulated in a form of lower-level algorithm. For the management level simulation the station is represented mainly by *time* elapsed between appropriate events (boarding/alighting time, time of waiting in a buffer etc.).

Other type of elementary nodes are capacitors or garages. Their internal structure is also not directly shown in the model. The only (user-defined) parameter of a garage is its capacity (the number of vehicles that can be stored).

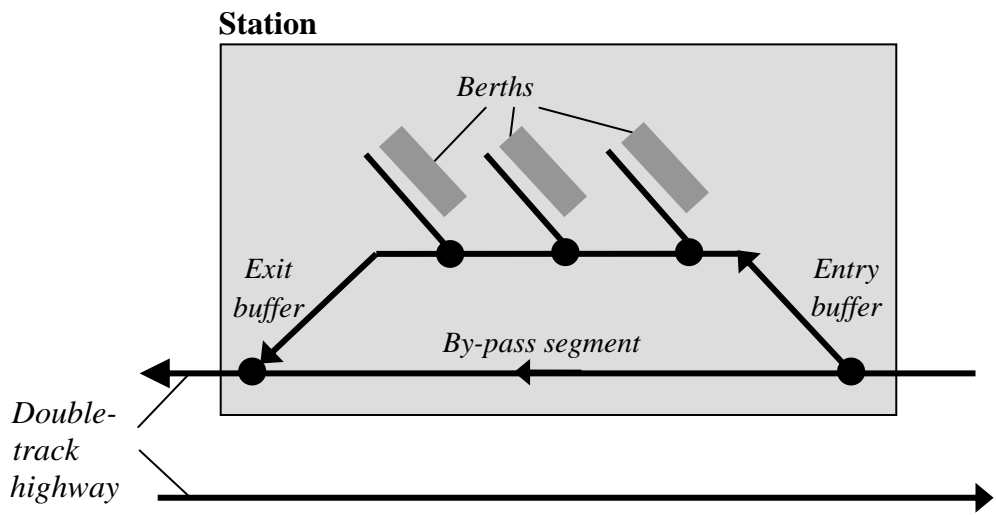

*Fig.* 2. Logical structure of a station

The four downtown stations are connected by an unidirectional track (the inner circle of 2 km diameter). The remaining part of the network is the two-directional (double track) highway (the outer circle of 4 km diameter and four radially diverging double tracks, 2 km each). In double-track highways the right-hand traffic is assumed. To shorten the distances among stations, double-track highways have also several U-turn connections located next to capacitors and stations F, H, J, L.

In order to avoid multilevel crossings all the crossings have the form of a roundabout (Fig. 3). Every roundabout involves a sequence of fork and join intersections connected with unidirectional track segments. Each individual segment is 20 meters long. Like in the conventional road traffic, vehicles already on roundabout have the priority over the incoming ones.

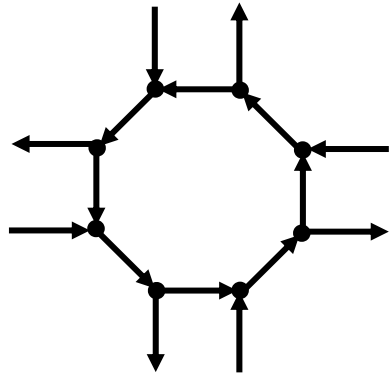

*Fig.* 3. Roundabout

Total length of tracks in *City* network (including roundabouts, tracks inside stations, U-turn connections etc.) is 52 000 m.

### 3.2.2. SeaShore

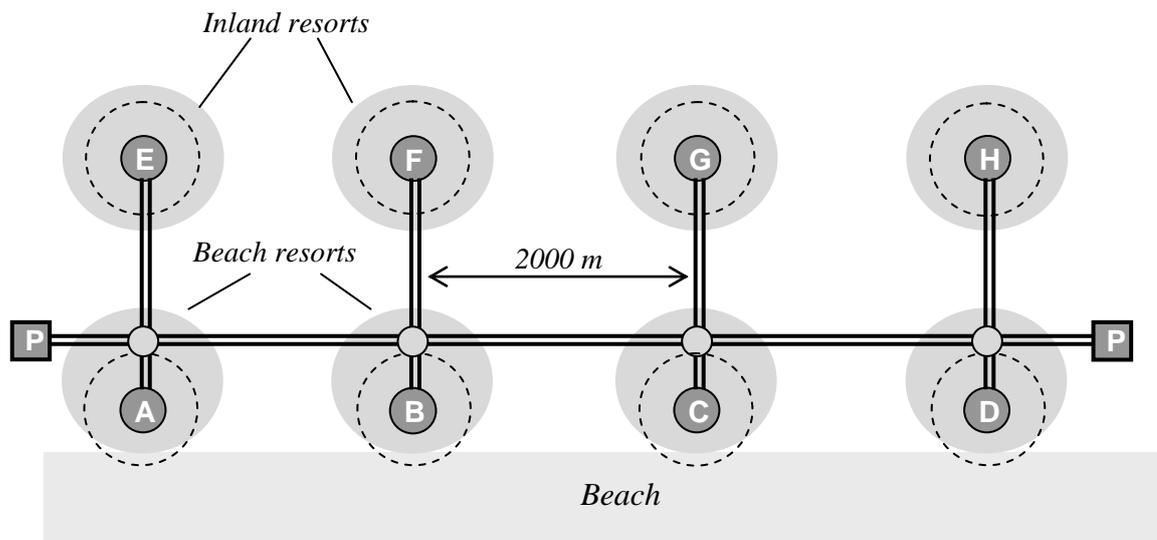

*Fig.* 4. *SeaShore* model

The *SeaShore* model (Fig. 4) corresponds to the linear topography of a hypothetical group of four seaside resorts located 2 km away from each other. They are connected by the double-track PRT highway that runs along the seafront boulevard. Each resort has a connection (also of the highway type) to its own additional hotels, spa etc. located 1 km inland. Two capacitors are placed in both ends of seafront highway. The schematic symbols used in Fig. 4 are the same as in Figure 1.

The total length of the track system of *SeaShore* network is 30 400 m.

### 3.2.3. TwinCity

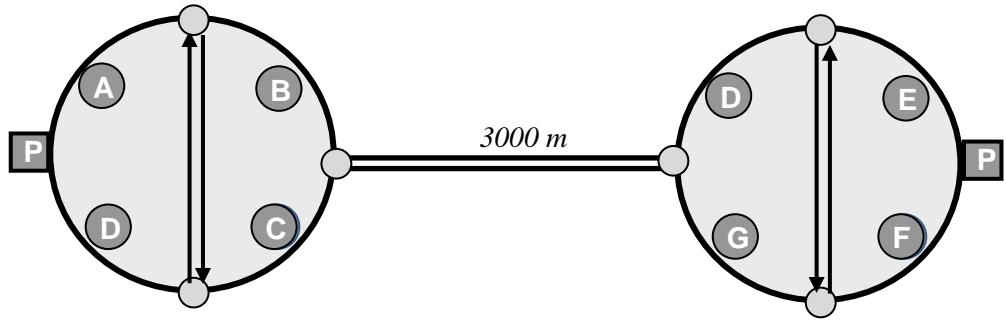

*Fig. 5. TwinCity* model

*TwinCity* model (Fig. 5) corresponds to a PRT network useful in a frequently occurring situation of two distant shopping centers, two parts of a university campus, two airport terminals etc.

Each of two centers is a round area of 1400 m diameter, framed by the single-track PRT highway. Inside the area there are four stations and one capacitor. Two opposite (the highest and the lowest) points of the area are additionally connected by two segments of ordinary track (i.e. with lower speed limit compared to the highway).

Two twin centers are connected by the double-track highway (3 000 m long, without any station on the way). The details of the network elements (stations, roundabouts etc.) as well as the schematic symbols used are the same as in the model of *City*. The total length of the track system in the *TwinCity* network is 23 000 m.

## 3.3. Common parameters and properties

Regardless of a different topography, all three models have a set of common assumptions, parameter values and properties.. Among most important are the following:
- Vehicles are of one type, with capacity of maximum 4 persons.
- Maximum speed on ordinary track is 10 m/s, on highway 15 m/s,
- Acceleration/deceleration is 2 m/s2,
- Separation between vehicles is 10 m,
- In each station there are: 5 berths, entry buffer for 5 vehicles, exit buffer for 5 vehicles.
- The boarding/alighting time is random, with the triangular distribution (10, 20, 30 seconds).
- The passengers arrive at stations in groups of 1 to 4 persons. The number of persons in a group is random with equal probabilities so that the mean cardinality of a group is 2,5 people.
- No ride sharing is allowed.
- The destination stations are selected random, with equal probabilities.

# 4. Estimating the demand and input rate

As the few existing networks [2, 14, 15] are in practice only the experimental solutions – the actual data on demand for PRT services are not available. We suggest that for the rough estimation of the demand the following procedure is applied.

Demand for transit services depends on how large is the population living in a given area. From the available statistical data we have estimated that for every thousand inhabitants, 20 people (i.e. 8.5 new groups of passengers) arrive to the network per hour. This value was chosen by the rough analogy to tram transport. As the passengers travel together in groups of 2.5 people (on average) – the above value means 8.5 new transit requests per hour per one thousand inhabitants.

We have assumed the population density equal to 3000 people/km2 for more densely populated areas (e.g. *City* center) and 2000 people/km2 for other areas (e.g. *SeaShore* resorts). Then, knowing the dimensions of a given area and the number of stations located in it we can compute how many transit requests arrive at every station per hour.

For isolated stations (as in *City* suburbs, *SeaShore* resorts etc.) it is assumed that they collect passengers from the area limited by the walking distance equal to 500 m.

For instance, the diameter of *City* downtown is 3 km, so its area is approximately 7 km2 and the downtown population is 21 000 inhabitants. Every hour 20*21=420 people arrive at the network grouped in 168 groups (2.5 people per group). So, the downtown population generates the demand of a total 168 vehicles per hour. Assuming that this input stream of passenger groups is equally distributed among four downtown stations – each station in the central zone should serve as much as 42 new passenger groups (i.e. 105 persons) per hour. Similarly, every isolated station (E to L in Fig. 1) collects the passengers from the walking zone area, i.e. from the circle of diameter equal to 1000 meters. The population living here (1570 people) generates 20*1.570=31.4 new passengers (i.e. 12.56 groups) per hour.

To sum up, in our model the whole *City* is populated by 33 560 inhabitants. This population generates the demand for approximately 168+8*12.5=268 vehicles per hour. The PRT network has to provide service to 268*2.5=570 people per hour. This value is the nominal *ridership* of *City* network.

Analogous procedure for estimating demand and the input rate has been applied also for other models.

For every *i*-th station, the estimated number of passenger groups per hour (denoted $\lambda_i$) characterizes the rate of the random input stream of customers (groups) demanding for the service. It is assumed that for every station the time between two consecutive arrivals is the exponentially distributed random variable with the parameter $\lambda_i$. In other words, during the simulation the time between two successive arrivals at *i*-th station is generated using the distribution function $F(t)=1-e^{-\lambda_i t}$. The mean inter-arrival time is $1/\lambda_i$.

So, the random input stream is the Poisson process. For $\lambda_i = const$ the process is *stationary*. Of course, in the practice all $\lambda_i$ are seldom constants. They may vary during the day, may depend on the day of the week etc. Nevertheless, the assumption that $\lambda_i$ 's are constants provides a good first approximation of the network traffic intensity.

Moreover, we suggest that in general the preliminary simulation experiments should be based on simplified rules, for instance that the input process is stationary, input rates are identical for all stations of the same type, the distribution of boarding time does not depend of the number of passengers in a vehicle, destination stations are selected random with equal probability and so on. Also, the network equilibrium should be preserved so that the total input rate is safely (e.g. twice) less than the maximum ridership for a given number of vehicles. Such assumptions, although they may seem unrealistic, provide a good reference for further investigation of more sophisticated variants of network properties.

# 5. Concluding remarks

Any benchmark should to some extent resemble real problems and solutions but in a simplified way so that they can be used independently of specific needs of individual application sites and of the simulation platform used. The above-described models fulfill this requirement but they are, of course, largely arbitrary. Any specific model feature or parameter can be challenged and disputed. What really matters is that the specification of such reference models should be determined, agreed upon and accepted by the research community. Also the new standard network structures (e.g., a grid model) are desirable.

Once established, the set of such reference models greatly facilitates the communication among the researchers and designers from a given field of science and engineering. The use of standardized benchmarks is the common practice in many branches. We can hope that it will prove very profitable also for the research on PRT networks.